\documentclass{PoS}
\usepackage{amsmath}

\title{Stable quark matter in cosmic rays ? }

\ShortTitle{Stable quark matter in cosmic rays ? }

\author{\speaker{Jes Madsen}\\
        Department of Physics and Astronomy, University of Aarhus,
        DK-8000 \AA rhus C, Denmark\\
        E-mail: \email{jesm@phys.au.dk}}

\abstract{
Stable lumps of quark matter may be present in cosmic rays at a flux level,
which can be detected by high precision cosmic ray experiments sensitive to
anomalous \textquotedblleft nuclei\textquotedblright\ with high
mass-to-charge ratio. The properties of these lumps, called strangelets, are
described, and so is the production and propagation of 
strangelets in cosmic rays. Two experiments underway which
are sensitive to a strangelet flux in the predicted range are briefly described.
Finally it is summarized how strangelets circumvent the acceleration problem
encountered by conventional candidates for ultra-high energy cosmic rays
and move the Greisen-Zatsepin-Kuzmin cutoff to energies well above the
observed maximum energies.
         }

\FullConference{29th Johns Hopkins Workshop on current problems in particle theory: Strong Matter in the Heavens\\
		 August 1--3, 2005\\
		 Budapest, Hungary}

\begin{document}

\section{Introduction}

Iron and nickel nuclei are normally assumed to be the most stable form of
hadronic matter at zero external pressure. In principle, this should be
testable from the basic theory of strong interactions, Quantum Chromo
Dynamics (QCD), but in practice this is impossible in the foreseeable future.
QCD is not suited for finite density calculations and calculations involving
many degrees of freedom, so much of our theoretical knowledge about dense
matter is based on phenomenological model calculations that try to
incorporate and parametrize some of the main features of the strong
interactions, such as confinement and asymptotic freedom. In some of these
studies, notably studies based on the MIT bag model, it has been shown, that
there is a significant range of parameters for which a three-flavor quark
phase with roughly equal numbers of up, down and strange quarks (called
strange quark matter in bulk, and strangelets in small lumps) has lower
energy than a nucleus with the same baryon number. Thus, strangelets rather
than nuclei may be the ground state of hadronic matter
\cite{Bodmer:1971we, Chin:1979yb, Witten:1984rs, Farhi:1984qu,
Madsen:1998uh, Weber:2004kj}.

A range of questions immediately occur if strangelets are more stable than
nuclei:

\begin{enumerate}
\item Why are we here---why don't our nuclei decay?

\item Can strangelets be created in the laboratory?

\item Can strange quark matter be found in space?
\end{enumerate}

The answers to all three questions rely on the properties of strangelets
summarized in the following. The answers to the questions (some of which
will be further explored below) are briefly:

\begin{enumerate}
\item Very low baryon number strangelets are likely to be unstable, even if
strange matter in bulk is absolutely stable, and for intermediate baryon
numbers the transformation of a nucleus requires an improbable high-order
weak interaction to simultaneously transform of order $A$ up and down quarks
into strange quarks, where $A$ is the baryon number. This explains the
stability of our nuclei. Strangelets have a positive electric charge, and
therefore repel nuclei from the surface, thus minimizing the risk of growth
via absorption of nuclei
\cite{Witten:1984rs, Farhi:1984qu, Dar:1999ac, Busza:1999pr, Madsen:2000kb,
Blaizot:2003bb}.

\item In principle strangelets might be formed in ultrarelativistic heavy-ion
collisions by coalescence or through a distillation mechanism leading to
strangeness enhancement. However, the available baryon number is low, which
makes it difficult to cross the low-$A$ stability cutoff, and furthermore
the environment is hot, so the process has rightly been compared to the
creation of ice cubes in a furnace. Experiments have been performed, but
with negative results
\cite{Sandweiss:2004bu, Klingenberg:2001qs, :2005cu}.

\item The most promising place to search for strange quark matter is in the
cosmos. Originally strange quark matter \textquotedblleft
nuggets\textquotedblright\ surviving from the cosmological quark-hadron
phase transition were suggested as a possible and even natural candidate for
the cosmological dark matter \cite{Witten:1984rs}, but later studies showed, 
that this was probably unlikely due to the high temperature 
environment that would lead to evaporation of the nuggets
\cite{Alcock:1985vc, Madsen:1986jg, Alcock:1988br, Olesen:1991zt,
Olesen:1993iz, Bhattacharjee:1993ah}. 
Recent lattice results indicating, that the
quark-hadron transition at low chemical potential (like the early Universe)
is not a first order phase transition seems to rule out the possibility of
cosmological strange quark matter. But strange quark matter, if stable, is
almost unavoidable in dense stellar objects. Pulsars and compact x-ray
sources will contain strange stars containing bulk strange quark matter
\cite{Witten:1984rs,Haensel:1986qb, Alcock:1986hz}, and
a number of observational signatures have been suggested to distinguish
these from compact stars made of ordinary hadronic matter. This is an
interesting story in its own right
(see \cite{Madsen:1998uh, Weber:2004kj} for reviews), 
but the focus of the present
presentation is the possibility of directly observing debris from collisions
of strange stars in the form of strangelets in cosmic rays.
\end{enumerate}

In the following I will summarize the main properties of strangelets, describe
the propagation of strangelets in cosmic rays, estimate the flux to be
expected in cosmic ray detectors, and discuss the possible detection in
a couple of future experiments. Finally I will argue that strangelets
may also be relevant for the puzzle of ultra-high energy cosmic
rays.

\section{Strangelet properties}

Neglecting quark Cooper pairing (to which we return in a moment), and
approximating the up and down quark current masses to zero, the mass of
strangelets in the MIT bag model \cite{Chodos:1974je} 
depends on three model parameters, namely
the strange quark mass, $m_{s}$, the bag constant, $B$, and the strong
fine-structure constant, $\alpha _{s}$. In most treatments, the strong
coupling and therefore $\alpha _{s}$ is set equal to zero. This is clearly
not correct at the relevant densities, but it has been shown
\cite{Farhi:1984qu}, that a nonzero
value can be mimicked very well by a scaling of the bag constant, and taking 
$\alpha _{s}=0$ simplifies calculations. A value of $B$ between $B^{1/4}=145
\mathrm{MeV}$ and $B^{1/4}=165\mathrm{MeV}$ allows bulk strange quark matter
to be absolutely stable for not-too high strange quark mass. In fact, for $
B^{1/4}<145\mathrm{MeV}$ even 2-flavor quark matter becomes absolutely
stable, having energy per baryon lower than $930\mathrm{MeV}$, which is
ruled out by the stability of nuclei. Such a transformation would not
require any weak conversion of quark flavor, whereas a similar decay into
strangelets requires almost one-third of the quarks transformed
simultaneously into strange quarks, a highly suppressed process.

Masses of strangelets can be calculated in the MIT bag model by explicitly
solving the Dirac equation with bag model boundary conditions, filling up,
down, and strange quark energy levels in the way which minimizes the mass
for a given $A$. Apart from details of closed shells the outcome of such
results can be reproduced fully in a multiple reflection expansion including
volume, surface, and curvature terms \cite{Farhi:1984qu, Berger:1986ps,
Gilson:1993zs, Madsen:1993ka, Madsen:1993iw, Madsen:1994vp}.

For a strange quark mass of $m_{s}=150\mathrm{MeV}$ and a bag constant $
B^{1/4}=145\mathrm{MeV}$, the mass of a strangelet in the lowest energy
ground state as a function of baryon number is given by
\cite{Madsen:1993iw, Madsen:1994vp}
\begin{equation}
M(A)=874\mathrm{MeV}A+77\mathrm{MeV}A^{2/3}+232\mathrm{MeV}A^{1/3}.
\end{equation}

Thus, for these parameters bulk quark matter is bound by $(930-874)\mathrm{
MeV}=56\mathrm{MeV}$ per baryon. 
Notice that the mass is composed of a bulk term
proportional to $A$, a surface term proportional to surface area or $A^{2/3}$
, and a non-negligible curvature term proportional to radius or $A^{1/3}$.
For this specific choice of parameters, only strangelets with baryon number $
A>23$ are stable relative to nuclei. Increasing the strange quark mass
and/or the bag constant moves the stability cut to higher $A$. Compared to
mass formulae for nuclei, the most striking features are the presence of a
significant curvature energy and the lack of a significant Coulomb energy.
Lack of the latter also means the absence of a minimum in $M/A$ as a
function of $A$.

The lack of a significant Coulomb energy is due to the fortuitous
cancellation of charge $+2e/3$ up quarks and charge $-e/3$ down and strange
quarks in strange quark matter with equal numbers of the three quark
flavors. Because of the non-zero s-quark mass the cancellation is not
perfect. Typical strangelets have slightly fewer strange quarks compared to
up and down, and therefore the net charge is slightly positive. A typical
model result (to be compared to $Z\approx 0.5A$ for nuclei) is
\cite{Heiselberg:1993dc}

\begin{align}
Z& =0.1\left( \frac{m_{s}}{150\mathrm{MeV}}\right) ^{2}A \\
Z& =8\left( \frac{m_{s}}{150\mathrm{MeV}}\right) ^{2}A^{1/3}
\end{align}
for $A\ll 700$ and $A\gg 700$ respectively (the slower growth for higher $A$
is a consequence of charge screening).

Thus a unique experimental signature of strangelets is an unusually high
mass-to-charge ratio compared to nuclei.

In recent years it has been shown, that quark matter at asymptotically high
density has an interesting property called color superconductivity
\cite{Alford:1997zt,Rapp:1997zu}. Even the
weakest attraction (and such attractions exist in QCD) leads quarks of
different colors and flavors to form pairs, much like Cooper pairs in a
superconductor, except that the binding in QCD is caused by a direct
attraction channel rather than via indirect phonon interaction. The binding
energy of a pair can be very large, ranging from MeV to over $100\mathrm{MeV}
$. In general these systems are called color superconductors. If all colors
and flavors pair in an equal manner one talks about color-flavor locking.

While the phenomenon of color-flavor locking seems generic in the infinite
density limit, the properties of strange quark matter at densities of order
or somewhat higher than nuclear matter density is at the focus of much
current research and discussion. This is the density regime of relevance for
strangelets, strange stars, and for quark matter cores in hybrid stars (the
analogs of neutron stars if quark matter is metastable so that it forms
above a certain density in compact star interiors). An additional binding
energy per baryon of $(10-100)\mathrm{MeV}$ is not unrealistic in these
systems, thus significantly increasing the likelihood of absolutely stable
strange quark matter and strangelets.

Since Cooper pairing involves quarks with equal (but opposite) momenta, the
natural ground state of a color-flavor locked system has equal Fermi momenta
for up, down, and strange quarks. Equal Fermi momenta implies equal number
densities, and therefore a total net quark charge of zero for a bulk system
\cite{Rajagopal:2000ff}.
However, a finite size strangelet has a surface suppression of massive
strange quarks relative to the almost massless ups and downs (massive
particle wave functions are suppressed at a surface), so that the total
charge of a color-flavor locked strangelet is positive and proportional to
area \cite{Madsen:2001fu}:

\begin{equation}
Z=0.3\left( \frac{m_{s}}{150\mathrm{MeV}}\right) A^{2/3}.
\end{equation}

This phenomenon persists even for very large bags, such as strange stars, so
color-flavor locked strange stars would also have a positive quark charge
\cite{Madsen:2001fu,Usov:2004iz,Stejner:2005mw}.

\section{Astrophysical strangelet production}

Strangelets may be produced when two
strange stars in a binary system collide due to loss of
orbital energy in the form of gravitational radiation. 
A strange star--black hole collision may also release lumps of quark
matter. If strange quark matter is absolutely stable
all compact stars are likely to be strange stars 
\cite{Madsen:1989pg,Friedman:1990qz}, and therefore the galactic coalescence
rate will be the one for double neutron star binaries recently updated in 
\cite{Kalogera:2003tn} based on available observations of binary pulsars to
be $83.0_{-66.1}^{+209.1}$ Myr$^{-1}$ at a 95\% confidence interval, thus of
order one collision in our Galaxy every 3,000--60,000 years
(but see \cite{deFreitasPacheco:2005ub} for a somewhat lower rate
estimate).

Each of these events involve a phase of tidal disruption of the stars as
they approach each other before the final collision. During this stage small
fractions of the total mass may be released from the binary system in the
form of strange quark matter. No realistic simulation of such a collision
involving two strange stars has been performed to date. Newtonian and
semirelativistic simulations of the inspiral of strange stars and black
holes do exist \cite{Lee:2002nk,Kluzniak:2002dm,Ratkovic:2005mn},
but the
physics is too different from the strange star-strange star collision to be
of much guidance. Simulations of binary neutron star collisions, depending on
orbital and other parameters, lead to the release of anywhere from $
10^{-5}-10^{-2}M_{\odot}$, where $M_{\odot}$ denotes the solar mass,
corresponding to a total mass release in the Galaxy of anywhere from $
10^{-10}-3\times 10^{-6}M_{\odot}$ per year with the collision rate above.
Given the high stiffness of the equation of state for strange quark matter,
strange star-strange star collisions should probably be expected to lie in
the low end of the mass release range, so the canonical input for the
following calculations is a galactic production rate of 
\begin{equation}
\overset{\cdot}{M}=10^{-10}M_{\odot}\text{yr}^{-1}.
\end{equation}

All strangelets released are assumed to have a single baryon number, $A$.
This is clearly a huge oversimplification, but there is no way of
calculating the actual mass spectrum to be expected. As demonstrated in \cite
{Madsen:2001bw} the quark matter lumps originally released by tidal forces
are macroscopic in size; when estimated from a balance between quark matter
surface tension and tidal forces a typical fragment size is
\begin{equation}
A\approx 4\times 10^{38}\sigma_{20}a_{30}^3,
\end{equation}
where $\sigma_{20}\approx 1$ is the surface tension in units of 20
MeV/fm$^2$ and $a_{30}$ is the distance between the stars in units of 30~km.
But subsequent collisions will lead to
fragmentation, and under the assumption that the collision energy is mainly
used to compensate for the extra surface energy involved in making smaller
strangelets, it was shown 
\cite{Madsen:2001bw} that a significant fraction of the mass released
from binary strange star collisions might ultimately be in the form of
strangelets with $A\approx 10^{2}-10^{4}$, though these values are strongly
parameter dependent.

The total flux results derived for cosmic ray
strangelets below are mostly such that values for some given $A$ are valid as a
lower limit for the flux for a fixed total strangelet mass injection if
strangelets are actually distributed with baryon numbers below $A$.

\section{Strangelet flux in cosmic rays
\protect\footnote{Section 4 closely follows Ref.~\cite{Madsen:2004vw}, 
but the treatment has been slightly simplified by excluding 
less important terms in the propagation equation.}}

Apart from an unusually high $A/Z$-ratio compared to nuclei, strangelets
behave in many ways like ordinary cosmic ray nuclei. For example, the
most likely acceleration mechanism would be Fermi acceleration in supernova
shocks resulting in a rigidity spectrum at the source which is a powerlaw in
rigidity as described below. Due to the high strangelet rigidity, $R$, at
fixed velocity, $v$ 
($R\equiv pc/Ze= Am_{0}c^{2}\gamma (\beta ) \beta/Ze$, where $
p$ is the strangelet momentum, $Am_0$ is the strangelet rest mass, 
$\beta\equiv v/c$, and $\gamma\equiv (1-
\beta^{2})^{-1/2}$) strangelets are more efficiently injected into an
accelerating shock than are nuclei with $A/Z\approx2$ (c.f. discussion of
nuclei in \cite{Gieseler}), and most strangelets passed
by a supernova shock will take part in Fermi acceleration.

The time scales for strangelet acceleration, energy loss, spallation and
escape from the Galaxy are all short compared to the age of the Milky Way
Galaxy. This makes it reasonable to assume that cosmic ray strangelets are
described by a steady state distribution given as a solution to a
propagation equation of the form 
\begin{equation}
\frac{dN}{dt}=0
\end{equation}
where $N(E,x,t)dE$ is the number density of strangelets at position $x$ and
time $t$ with energy in the range $[E,E+dE]$.

Given a solution for $N(E)$ the corresponding flux in the \textquotedblleft
average\textquotedblright\ interstellar medium with energies from $[E,E+dE]$
is given by
\begin{equation}
F_{E}(E)dE=\frac{\beta c}{4\pi}N(E)dE\text{,}
\end{equation}
and the corresponding flux in terms of rigidity is
\begin{equation}
F_{R}(R)dR=Ze\beta F_{E}(E(R))dR
\end{equation}
(using $dE/dR=Ze\beta$).

Like other charged cosmic ray particles strangelets are influenced by the
solar wind when entering the inner parts of the Solar System. The detailed
interactions are complicated, but as demonstrated for nuclei in \cite
{Gleeson:1968}, a good fit to the solar modulation of the cosmic ray
spectrum can be given in terms of a potential model, where the charged
particle climbs an electrostatic potential of order $\Phi=500$ MeV (the
value changes by a factor of less than 2 during the 11 (22) year solar
cycle). This effectively reduces the cosmic ray energy by $|Z|\Phi$ relative
to the value in interstellar space, and at the same time the flux is reduced
by the relative reduction in particle momentum squared, so that the
modulated spectrum is
\begin{equation}
F_{\text{mod}}(E)=\left( \frac{R(E)}{R(E+|Z|\Phi)}\right)
^{2}F_{E}(E+|Z|\Phi)\text{.}
\end{equation}
Solar modulation significantly suppresses the flux of charged cosmic rays at
energies below a few GeV and effectively works like a
smooth cutoff in flux below kinetic energy of order $|Z|\Phi$. Since
strangelets have a high mass-to-charge ratio they are nonrelativistic at
these energies, which correspond to rigidities of $R_{\text{SM}
}\approx(A/Z)^{1/2}\Phi_{500}^{1/2}~GV$, where $\Phi_{500}=\Phi/(500$MeV$).$

For cosmic rays to reach the Earth or an Earth-orbiting detector like the
Alpha Magnetic Spectrometer on the International Space Station, the rigidities
have to exceed the geomagnetic cutoff rigidity, which is a function of
detector position, and for an orbiting observatory like AMS the value varies
from 1--15 GV as a function of time. The geomagnetic cutoff
rigidity for low mass strangelets is comparable to or higher than the solar
modulation cutoff, whereas high mass strangelets experience solar modulation
already at rigidities above the geomagnetic cutoff.

For a non-magnetic body like the Moon, there is no corresponding cutoff, and
the total flux is given by $F_{\text{mod}}$. This makes the lunar
surface an interesting laboratory for a strangelet search (c.f.\ 
Section~\ref{moon}; see also Ref.~\cite{Monreal:2005dg}).

Given the significant uncertainty in input parameters a
simple but physically transparent model for strangelet propagation was
chosen in \cite{Madsen:2004vw}.
$\frac{dN}{dt}$ is given by the following sum of a source term from
supernova acceleration, a diffusion term, loss terms due to escape from the
Galaxy, energy loss, and spallation,
\begin{equation}
\frac{dN}{dt}=\frac{\partial N}{\partial t}|_{\text{source}}+D\nabla ^{2}N+
\frac{\partial N}{\partial t}|_{\text{escape}}+\frac{\partial }{\partial E}
[b(E)N]+\frac{\partial N}{\partial t}|_{\text{spallation}}.
\end{equation}
The individual terms will be defined and discussed in the following.
Further terms describing, e.g. decay and reacceleration of strangelets
due to passage of new supernova shock waves, are discussed in 
\cite{Madsen:2004vw}, but will be neglected here for simplicity.

The strangelet spectrum after acceleration in supernova shocks is assumed to
be a standard powerlaw in rigidity with index $-2.2$ as derived from
observations of ordinary cosmic rays. The minimal rigidity is assumed to be
given by the speed of a typical supernova shock wave, $\beta _{SN}\approx
0.005,$ so $R_{\min }=\gamma (\beta
_{SN})\beta _{SN}Am_{0}c^{2}/Ze\approx 5$MV$A/Z$. The maximal rigidity from
acceleration in supernova shocks, $R_{\max },$ is of order $10^{6}$ GV, but
the actual number is irrelevant since $R_{\max }\gg R_{\min }$ and the
rigidity spectrum steeply declining. For a total
production rate of $\overset{\cdot }{M}=10^{-10}M_{\odot }$yr$^{-1}$ of
baryon number $A$ strangelets spread evenly in an effective galactic volume
$V$, the total source term is

\begin{equation}
G(R)=1.2\frac{\overset{\cdot }{M}}{VAm_{0}R_{\min }}\left( \frac{R}{R_{\min }
}\right) ^{-2.2}
\end{equation}
or in terms of energy (using $dE/dR=Ze\beta $ and $G(R)dR=G(E)dE$)
\begin{equation}
\frac{\partial N}{\partial t}|_{\text{source}}\equiv G(E)=\frac{G(R(E))}{
Ze\beta }.
\end{equation}

The terms $D\nabla ^{2}N+\frac{\partial N}{\partial t}|_{\text{escape}}$,
where $D$ is the diffusion coefficient, describe cosmic ray diffusion in
real space and eventual escape from the confining magnetic field of the
Galaxy. Charged cosmic rays are spiralling along field lines in the weak
galactic magnetic field, but due to the very irregular structure of the
field, the particles scatter on magnetic \textquotedblleft
impurities\textquotedblright , and the motion is best described in terms of
diffusion. From studies of nuclei it is known, that cosmic rays are confined
to move in a region significantly larger than the galactic disk, where most
stars and interstellar matter is located. A typical value for the
effective galactic volume confining cosmic rays is $V=1000$ kpc$^{3}$.
The standard leaky box approximation assumes $D=0$
and $\frac{\partial N}{\partial t}|_{\text{escape}}=-N/\tau _{\text{escape}}$,
where $\tau _{\text{escape}}(A,Z,E)$ is the average escape time from an
otherwise homogeneous distribution in the galactic volume, $V$. From studies
of cosmic ray nuclei the escape time is known as a function of rigidity, $R$
, as
\begin{equation}
\tau _{\text{escape}}=\frac{8.09\times 10^{6}\text{y}}{n\beta }\left( \frac{R
}{R_{0}}\right) ^{\delta }.
\end{equation}
where $R_{0}=4.7$GV, $\delta =0.8$ for $R<R_{0}$, and $\delta =-0.6$ for $
R>R_{0}$. $n$ denotes the average hydrogen number density per cubic
centimeter ($n\approx 0.5$ when averaging over denser regions in the
galactic plane and dilute regions in the magnetic halo).

The term in the propagation equation $\frac{\partial }{\partial E}[b(E)N]$,
describes the influence of energy loss processes. The energy loss rate $
b(E)\equiv -dE/dt$ can be treated as a sum of ionization losses (from
interaction with neutral hydrogen atoms and molecules), Coulomb losses (from
interaction with ionized hydrogen), and pion production losses from
inelastic collisions at high relativistic $\gamma $-factor (threshold at $
\gamma =1.3$). The various contributions
are described in \cite{Madsen:2004vw}. At speeds close to the
speed of light the ionization loss is simply proportional to $nZ^{2}$.

Like nuclei strangelets have a roughly geometrical cross section
proportional to $A^{2/3}$ for spallation in collisions with interstellar
matter (mainly hydrogen). The corresponding spallation time scale is taken
to be
\begin{equation}
\tau _{\text{spallation}}=\frac{2\times 10^{7}\text{y}}{n\beta }A^{-2/3}.
\end{equation}
At low kinetic energy the cross section can vary somewhat due to
resonances etc, but such complications will be neglected here, since the
detailed physics is unknown in the case of strangelets. We have also
neglected the slight reduction in geometrical area of strangelets relative
to nuclei due to their slightly larger density. The largest uncertainty in
the treatment of spallation is the fact that strangelets (like nuclei) are
not always completely destroyed in a spallation reaction. In addition to
nucleons and nuclei smaller strangelets may result from this type of
reaction, but we are ignorant of the physics to an extent where it is
impossible to include this effective feed-down to lower $A$ in a meaningful
manner. Therefore, spallation is assumed to be a process destroying
strangelets, i.e.
\begin{equation}
\frac{\partial N}{\partial t}|_{\text{spallation}}=-\frac{N}{\tau _{\text{
spallation}}}.
\end{equation}
This leads to an overall underestimate of the strangelet flux.

Strangelet energies are redistributed according to the propagation equation.
Some leave the Galaxy or are destroyed by spallation. Occasionally
strangelets get a new kick from a passing supernova shock, and in a first
approximation they regain the source term relative distribution of rigidity.
The time scale between supernova shock waves passing a given
position in interstellar space is of order $\tau _{SN}\approx 10^{7}$ y.
This scale is comparable to or larger than the time scales for energy loss,
spallation, and escape from the Galaxy, so reacceleration of cosmic ray
strangelets has only a moderate influence on the steady state distribution.
By adding energy (on average) to the particles it actually reduces the total
flux of strangelets somewhat because higher energies make destruction and
escape more likely (see Ref.~\cite{Madsen:2004vw} for details).

Introducing the terms discussed above the steady state equation 
$dN/dt=0$ leads to the following differential equation for $N(E)$
\begin{equation}
b(E)\frac{dN}{dE}=\frac{N(E)}{\tau (E,A,Z)}-G(E),
\end{equation}
where
\begin{equation}
1/\tau (E,A,Z)=1/\tau _{\text{escape}}+1/\tau _{\text{spallation}}+1/\tau _{
\text{loss}},
\end{equation}
with $1/\tau _{\text{loss}}\equiv -db(E)/dE.$ When energy loss can be
neglected, $b(E)\approx 0$ and $|\tau _{\text{loss}}|\approx \infty $. In
this limit the spectrum is simply given by
\begin{equation*}
N(E)\approx G(E)\tau (E,A,Z),
\end{equation*}
and $1/\tau (E,A,Z)\approx 1/\tau _{\text{escape}}+1/\tau _{\text{spallation}
}$.

The general solution of the propagation equation requires numerical
integration, but several limits can be treated analytically and provide a
physical understanding of the full numerical solutions 
found in \cite{Madsen:2004vw}. The special cases
(disregarding solar modulation and geomagnetic cutoff) can be divided
according to the relative importance of the different time scales, $\tau _{
\text{escape}}$, $\tau _{\text{spallation}}$, and $|\tau _{\text{loss}}|$
(the energy loss time scale is negative at low and high energies, describing
a net increase in number of particles).

When one of the time scales is significantly smaller than the others at a
given energy (rigidity), the corresponding process dominates the physics.
The relative importance of the processes depends on strangelet properties $
A,Z$, on the density of interstellar hydrogen $n$ (though most processes
have the same $n$-dependence), and of course on the strangelet energy, $E$
(or rigidity, $R$). Energy loss dominates at low energy, spallation at
intermediate $E$, and escape from the Galaxy at the highest energies.

The energy loss domination is important only at energies 
below the solar modulation cutoff, so I refer the reader to 
\cite{Madsen:2004vw} for details.

At intermediate energies the spectrum is determined by the strangelet
spallation time, $\tau \approx \tau _{\text{spallation}}$, and the
energy distribution is approximately given by
\begin{equation}
N(E)\approx G(E)\tau _{\text{spallation}}(n,\beta ,A)
\end{equation}
with $\tau _{\text{spallation}}=2\times 10^{7}$y$n^{-1}\beta ^{-1}A^{-2/3}.$

This gives the approximate result
\begin{equation}
F_{R}(R)=2.34\times 10^{5}\text{m}^{-2}\text{yr}^{-1}\text{sterad}^{-1}\text{
GV}^{-1}A^{-0.467}Z^{-1.2}R_{\text{GV}}^{-2.2}\Lambda ,
\end{equation}
or for the total flux above rigidity $R$ 
\begin{equation}
F(>R)=1.95\times 10^{5}\text{m}^{-2}\text{yr}^{-1}\text{sterad}
^{-1}A^{-0.467}Z^{-1.2}R_{\text{GV}}^{-1.2}\Lambda .
\end{equation}

The results scale in proportion to 
\begin{equation}
\Lambda =\left( \frac{\beta _{\text{SN}}}{0.005}\right) ^{1.2}\left( \frac{
0.5\text{cm}^{-3}}{n}\right) \left( \frac{\overset{\cdot }{M}}{
10^{-10}M_{\odot }\text{yr}^{-1}}\right) \left( \frac{1000\text{kpc}^{3}}{V}
\right) \left( \frac{930\text{MeV}}{m_{0}c^{2}}\right) .  \label{lambda}
\end{equation}
Notice that the differential strangelet spectrum keeps the source term
slope, $G(R)\propto R^{-2.2}$.

The intermediate energy domain is replaced by the high energy domain when $
\tau _{\text{escape}}\leq \tau _{\text{spallation}}$. Except for very low $A$
this happens when $R>1.0$GV$A^{1.11}$, or $E>1.0$GeV$ZA^{1.11}\beta ^{-1}$.
At high energies the spectrum is determined by the confinement time of
strangelets in the Galaxy, $\tau \approx \tau _{\text{escape}}$, and the
propagation equation leads to
\begin{equation}
N(E)\approx G(E)\tau _{\text{escape}}(n,\beta ,R).
\end{equation}
For semirelativistic or relativistic strangelets with $\beta \approx 1$, $
\tau _{\text{escape}}\propto R^{-0.6}$, so the spectrum is steepened from
the source term $R^{-2.2}$ to $R^{-2.8}$.

This gives the approximate result
\begin{equation}
F_{R}(R)=2.40\times 10^{5}\text{m}^{-2}\text{yr}^{-1}\text{sterad}^{-1}\text{
GV}^{-1}A^{0.2}Z^{-1.2}R_{\text{GV}}^{-2.8}\Lambda ,
\end{equation}
and for the total flux above rigidity $R$ 
\begin{equation}
F(>R)=1.33\times 10^{5}\text{m}^{-2}\text{yr}^{-1}\text{sterad}
^{-1}A^{0.2}Z^{-1.2}R_{\text{GV}}^{-1.8}\Lambda .
\end{equation}

The astrophysical input parameters in the present calculations are
uncertain at the order of magnitude level, so approximate 
relations for the total flux of
strangelets hitting the Earth or Moon accurate within a factor of $2$ (for
fixed input parameters) are useful. As indicated above solar modulation
effectively cuts off the strangelet flux at rigidities of order $R_{\text{SM}
}\approx(A/Z)^{1/2}\Phi_{500}^{1/2}$GV, which is in the part of the spectrum
where the strangelet flux is governed by spallation. The total flux hitting
the Moon or Earth is therefore roughly given by
\begin{equation}
F_{\text{total}}\approx2\times10^{5}\text{m}^{-2}\text{yr}^{-1}\text{sterad}
^{-1}A^{-0.467}Z^{-1.2}\max[R_{\text{SM}},R_{\text{GC}}]^{-1.2}\Lambda,
\end{equation}
depending on whether solar modulation or geomagnetic cutoff dominates. In
the case of solar modulation domination (always relevant for the Moon, and
relevant for AMS as long as $R_{\text{SM}}>R_{\text{GC}}$) one obtains
\begin{equation}
F_{\text{total}}\approx2\times10^{5}\text{m}^{-2}\text{yr}^{-1}\text{sterad}
^{-1}A^{-1.067}Z^{-0.6}\Phi_{500}^{-0.6}\Lambda.  \label{appflux}
\end{equation}
For strangelets obeying the CFL mass-charge relation $Z=0.3A^{2/3}$ this
becomes 
\begin{align}
F_{\text{total}}\approx~& 4\times10^{5}\text{m}^{-2}\text{yr}^{-1}\text{
sterad} ^{-1}A^{-1.467}\Phi_{500}^{-0.6}\Lambda \\
\approx~& 2.8\times10^{4}\text{m}^{-2} \text{yr}^{-1}\text{sterad}
^{-1}Z^{-2.2}\Phi_{500}^{-0.6}\Lambda,
\end{align}
which reproduces the numerical results of Ref.~\cite{Madsen:2004vw}
to within 20\% for $Z>10$ and to
within a factor of a few even for small $Z$, where the assumptions of
nonrelativistic strangelets and spallation domination both are at the limit
of being valid.

The strangelet flux for $\Lambda$ of order unity is high enough
to be of interest for various upcoming experimental searches, and at the
same time small enough to agree with previous searches which have given
upper limits or shown marginal evidence for signatures consistent with
strangelets (see \cite{Sandweiss:2004bu} for an overview). As stressed
several times above, many parameters are uncertain at the order of magnitude
level. The scaling with these parameters is indicated where relevant. In
particular this is true for the overall normalization of the strangelet flux
as expressed via the parameter $\Lambda$ (Eq.\ (\ref{lambda})), whereas the
relative behavior of the differential flux is less uncertain.

Apart from the uncertainty in parameters within the picture discussed here,
one cannot rule out the possibility that some of the basic assumptions need
to be modified. In addition to strangelet production in strange star
collisions, it has been suggested that a (possibly small) flux of
strangelets may be a direct outcome of the Type II supernova explosions \cite
{Benvenuto:1989hw}, where strange stars form.
The treatment above does not include such additional
strangelet production mechanisms, and therefore the flux predictions are
conservative. Another assumption that leads to a conservative lower limit on
the flux is that spallation is assumed to destroy strangelets completely. At
least for low-energy collisions one would expect that fragments of
strangelets would survive, but the input physics for performing a realistic
strangelet spallation study is not sufficiently well known, and therefore
the conservative assumption of complete destruction was made.
A numerical simulation of strangelet propagation in Ref.~\cite
{Medina-Tanco:1996je} assumed stripping of nucleons rather than complete
destruction of strangelets in interstellar collisions and studied two
specific sets of values for the mass and energy of strangelets injected into the
interstellar medium. Several other assumptions made in that numerical study
differ from those of the current investigation, so a direct
comparison of the results is not possible, except
that both studies are consistent with the possibility of a significant,
measurable strangelet flux in our part of the Galaxy.

But ultimately the question of whether strangelets exist in cosmic rays is
an experimental issue.

\section{Experiments underway}

Several experiments have searched for strangelets in cosmic rays and/or have
had their data reanalyzed to provide limits on the strangelet abundance.
While some interesting events have been found that are consistent with the
predictions for strangelets, none of these have been claimed as real
discoveries. Interpreted as flux limits rather than detections these results
are consistent with the flux estimates given above. For discussions see 
\cite{Sandweiss:2004bu}.

If the interesting events were actual measurements, two experiments that are
currently underway (with the author as a humble theoretician participating
in collaborations with very capable experimentalists!) 
will reach sensitivities, that
would provide real statistics. These experiments are the Alpha Magnetic
Spectrometer (AMS-02) on the International Space Station, and the Lunar Soil
Strangelet Search (LSSS) at the Wright Nuclear Structure Laboratory at Yale.

\subsection{AMS-02}

The Alpha Magnetic Spectrometer (AMS) is a high profile space-based particle
physics experiment involving approximately 500 physicists from more than 50
institutions in 16 countries, led by Nobel laureate Samuel Ting of the
Massachusetts Institute of Technology (MIT). The AMS flew in space in June
of 1998 aboard the Space Shuttle Discovery
\cite{Aguilar:2002ad}, and it is currently scheduled to
fly to the International Space Station (ISS) on Utilization Flight \# 4.1
(UF4.1) with launch in 2008. Once on the ISS, the AMS-02 will remain active
for at least three years before it is returned to Earth. AMS-02 will provide
data with unprecedented accuracy on cosmic ray electrons, positrons,
protons, nuclei, anti-nuclei and gammas in the GV-TV range in order to probe
issues such as antimatter, dark matter, cosmic ray formation and
propagation. And in addition it will be uniquely suited to discover
strangelets characterized by extreme rigidities for a given velocity
compared to nuclei \cite{Sandweiss:2004bu}. 
AMS-02 will have excellent charge resolution up to $
Z\approx 30$, and should therefore be able to probe a large mass range for
strangelets. A reanalysis of data from the 1998 AMS-01 mission has given
hints of some interesting events, such as one with $Z=2,A=16$
\cite{Choutko:2003}, but with the
larger AMS-02 detector running for 3 years or more, real statistics is
achievable.

\subsection{LSSS}
\label{moon}

The Lunar Soil Strangelet Search (LSSS) is a search for $Z=8$ strangelets
using the tandem accelerator at the Wright Nuclear Structure Laboratory at
Yale. The experiment involves a dozen people from Yale, MIT, and \AA rhus,
led by Jack Sandweiss of Yale. The experiment which is currently in its
calibration phase, studies a sample of 15 grams of lunar soil from Apollo 11
(sample number AH10084). The expectation is to reach a sensitivity of $
10^{-17}$ over a wide mass range, sufficient to provide detection according
to Equation~\eqref{appflux} 
if strangelets have been trapped in the lunar surface layer,
which has an effective cosmic ray exposure time of around 500 million years.

\section{Moving the GZK-cutoff--Strangelets as ultra-high energy cosmic rays}

One of the most interesting puzzles in cosmic ray physics is the apparent
existence of cosmic rays with energies well beyond $10^{19}\mathrm{eV}$,
with measured energies as high as $3\times 10^{20}\mathrm{eV}$. Briefly
speaking there are two puzzles \cite{Greisen:1966jv,Zatsepin:1966jv}:

\begin{enumerate}
\item It is almost impossible to find an astrophysical mechanism capable of
accelerating cosmic rays to these energies.

\item Even if acceleration happens, ultra-high energy cosmic rays have a
relatively short mean-free path for interactions with the cosmological
microwave background photons, and only cosmic rays from fairly nearby
(unknown) sources would be able to reach us with the high energies measured.
\end{enumerate}

Interestingly, strangelets circumvent both of these problems, and therefore
provide a possible mechanism for cosmic rays beyond the socalled
Greisen-Zatsepin-Kuzmin (GZK) cutoff. The benefits of strangelets with
respect to the two problems are \cite{Madsen:2002iw}:

\begin{enumerate}
\item All astrophysical cosmic ray \textquotedblleft
accelerators\textquotedblright\ involve electromagnetic fields, and
the maximal energy that can be transmitted to a charged particle is
proportional to the charge, $Z$. For instance some mechanisms can provide a
maximal rigidity, $R_{\mathrm{MAX}}$, proportional to the product of the
magnetic field and size of the region where acceleration takes place, and
for relativistic particles this corresponds to a maximal energy $E_{\mathrm{
MAX}}=ZR_{\mathrm{MAX}}$. The charge of massive strangelets has no upper
bound in contrast to nuclei, so highly charged strangelets are capable of
reaching energies much higher than those of cosmic ray protons or nuclei
using the same ``accelerator'' \cite{Madsen:2002iw}.

\item The GZK-cutoff is a consequence of ultrarelativistic cosmic ray
projectiles hitting a $2.7\mathrm{K}$ background photon with a
Lorentz-factor $\gamma $ large enough to boost the $7\times 10^{-4}\mathrm{eV
}$ photon to energies beyond the threshold of a process leading to
significant projectile energy loss, such as photo-pion production,
photo-pair production, or photo-disintegration. The threshold for such a
process has a fixed energy, $E_{\mathrm{Threshold}}$, in the frame of the
cosmic ray (e.g., $E_{\mathrm{Threshold}}\approx 10\mathrm{MeV}$ for
photo-disintegration of a nucleus or a strangelet), corresponding to a
Lorentz-factor $\gamma _{\mathrm{Threshold}}=E_{\mathrm{Threshold}}/E_{2.7
\mathrm{K}}$ ($\approx 10^{10}$ for photo-disintegration). 
But this corresponds to a cosmic ray total energy
\begin{equation}
E_{\mathrm{Total}}=\gamma _{\mathrm{Threshold}}Am_{0}c^2.
\end{equation}
Thus, $E_{\mathrm{Total}}$ ($\approx 10^{19}A~\mathrm{eV}$ at the
photo-disintegration threshold) is proportional
to the rest mass of the cosmic ray, and since strangelets can have much
higher $A$-values than nuclei, this pushes the GZK-cutoff energy to values
well beyond the current observational limits for 
ultra-high energy cosmic rays \cite{Madsen:2002iw,Rybczynski:2001bw}.
\end{enumerate}

A testable prediction of the strangelet scenario for ultra-high energy
cosmic rays is that strangelets at a given energy will be more isotropically
distributed than protons and nuclei because the gyro-radius in the galactic
magnetic field is proportional to $Z^{-1}$, so the arrival direction for
high-$Z$ strangelets points back to the source to a lesser extent than for
low-$Z$ candidates which have gyro-radii comparable to the size of the
Galaxy.

\section{Conclusion}

The total
strangelet flux reaching the Moon or a detector in Earth orbit is in a
regime that could be within experimental reach, and therefore provide a
crucial test of the hypothesis of absolutely stable strange quark matter.
Experiments are underway which are sensitive to the high mass-to-charge
signature expected for such events within the flux-range predicted.
Strangelets may also provide an interesting explanation of ultra-high
energy cosmic ray events.

\acknowledgments
This work was supported by the Danish Natural Science Research Council.

\end{document}